\documentclass{elsart}

\def\cs{$^{137}$Cs~}
\def\ces{$^{134}$Cs~}
\def\rb{$^{87}$Rb~}

\def\naitl{NaI(T$\ell$)~}
\def\csitl{CsI(T$\ell$)~}
\def\PREP{Phys. Rep.}

\def\Journal#1#2#3#4{{#1} {\bf #2} (#3) #4}

\def\NIMA{Nucl. Instrum. Methods A}

\def\PLB{Phys. Lett.  B}
\def\PRL{Phys. Rev. Lett.}
\def\PRD{Phys. Rev. D}

\def\ASP{Astropart. Phys.}
\def\JKPS{J. Kor. Phys. Soc.}
\def\etal{{\it et al.}}

\usepackage{amssymb}
\usepackage{epsfig}

\begin{document}

\begin{frontmatter}

\title{First limit on WIMP cross section with low background \csitl crystal detector}

\author[snu]{H.S.~Lee\corauthref{hslee}},
 \author[snu]{H.~Bhang},
 \author[snu]{J.H.~Choi},
 \author[ewha]{I.S.~Hahn},
 \author[tsinghua]{D.~He},
 \author[yonsei]{M.J.~Hwang},
 \author[knu]{H.J.~Kim},
 \author[snu]{S.C.~Kim},
 \author[snu]{S.K.~Kim\corauthref{skkim}},
 \author[snu]{S.Y.~Kim},
 \author[snu]{T.Y.~Kim},
 \author[sejong]{Y.D.~Kim},
 \author[snu]{J.W.~Kwak},
 \author[yonsei]{Y.J.~Kwon},
 \author[snu]{J.~Lee},
 \author[snu]{J.H.~Lee},
 \author[sejong]{J.I.~Lee},
 \author[snu]{M.J.~Lee},
 \author[tsinghua,ihep]{J.~Li}
 \author[snu]{S.S.~Myung},
 \author[snu]{H.~Park\thanksref{hpark}},
 \author[snu]{H.Y.~Yang},
 \author[tsinghua]{J.J.~Zhu},
\center (\author{KIMS Collaboration})

\corauth[hslee]{tgsh@hep1.snu.ac.kr}
\corauth[skkim]{skkim@hep1.snu.ac.kr}
\thanks[hpark]{Present address : Division of Chemical Metrology and
Materials Evaluation, Korea Research Institute of Standards and
Science, Yuseong, Daejeon, 305-600, Republic of Korea}

\address[snu]{DMRC and School of Physics, Seoul National University, Seoul 151-742, Korea}
\address[sejong]{Department of Physics, Sejong University, Seoul 143-747, Korea}
\address[yonsei]{Physics Department, Yonsei University, Seoul 120-749, Korea}
\address[ewha]{Department of Science Education, Ewha Womans
University, Seoul 120-750, Korea}
\address[knu]{Physics Department, Kyungpook National University, Daegu
702-701, Korea}
\address[tsinghua]{Department of Engineering Physics, Tsinghua University,
Beijing 100084, China}
\address[ihep]{Institute of High Energy Physics, Chinese Academy of Sciences,
Beijing 100039, China}

\begin{abstract}
The Korea Invisible Mass Search~(KIMS) collaboration has been carrying
out WIMP search experiment with \csitl crystal detectors at the YangYang
Underground Laboratory. A successful reduction of the internal background
of the crystal was done and a good pulse shape discrimination was
achieved. We report the first result on WIMP search obtained with
237~kg$\cdot$days data using one full-size \csitl crystal of 6.6~kg mass.

\end{abstract}

\end{frontmatter}

\section{Introduction}

Although the existence of dark matter as a major portion of the
matter in the universe has been well supported by various
astronomical observations, its identity is unknown yet. Weakly
Interacting Massive Particle~(WIMP) is regarded as one of the
strongest candidates for cold dark matter particle~\cite{Jungman}
and many experimental searches for WIMPs have been performed. An
indication of an annual modulation of the signal reported by DAMA using
\naitl crystal detectors may be a possible evidence of WIMP
signal~\cite{DAMA}. However, stringent limits set by cryogenic
detectors, CDMS~\cite{CDMS} and EDELWEISS~\cite{EDELWEISS} seem to rule
out the DAMA signal region. Still,
ways of interpreting both results without
conflict are not completely excluded because of the difference in
experimental techniques and target
nuclei~\cite{dama7year,reconcil,spindep}.

The Korea Invisible Mass Search~(KIMS) collaboration has been carrying
out the WIMP search with \csitl crystals. Low threshold suitability
for WIMP search and pulse shape discrimination~(PSD) superiority to
\naitl has been
demonstrated~\cite{kims_hjkim,kims_hspark,pecourt,kudryavtsev}. Even
if the \csitl crystal has the advantage of a good PSD and the ease of
getting a large detector mass, the internal background including \cs,
\ces and, \rb has been a major hurdle to apply the crystal to WIMP
search~\cite{kudryavtsev,kims_tykim}. We have successfully purified
\csitl crystals after extensive studies on the contamination
mechanism. A pilot experiment with a low background \csitl crystal of
6.6~kg mass has been carried out at the YangYang Underground
Laboratory~(Y2L) in Korea. We report the first result obtained with
237~kg$\cdot$days of data taken with one crystal of 8~x~8~x~23~cm$^3$
. 

\section{YangYang Underground Laboratory}

We have established an underground laboratory at YangYang
utilizing the space provided by the YangYang Pumped Storage Power
Plant currently under construction by Korea Midland Power Co.
The underground laboratory is located in a tunnel where the
vertical earth overburden is approximately 700~m. The muon flux measured
with the muon detector is $2.7\times10^{-7}/\rm{cm}^{2}/s$, which
is consistent with the water equivalent depth of
2000~m~\cite{kims_zhu}. The
laboratory is equipped with a clean room with an air conditioning
system for a constant temperature and low humidity. An environment
monitoring system is installed for continuous monitoring
of temperature and humidity. The temperature inside the \csitl
detector container is stable within $\pm$~0.2~$^o$C. The rock
composition surrounding the laboratory was analyzed with the ICP-MASS method
and the contamination of
$^{238}$U and $^{232}$Th is reported to be at a level of $<$~0.5~ppm 
and 5.6~$\pm$~2.6~ppm respectively. The relatively low
contamination of $^{238}$U in the rocks of the tunnel results in a low
level of radon contamination in the air of the tunnel. A radon
detector~\cite{kims_radon} was constructed to monitor the level of
radon in the experimental hall. The contamination level of
$^{222}$Rn in the tunnel air was 1-2~pCi/$\ell$ which is
slightly lower than other underground laboratories such as Gran
Sasso~\cite{rn_gran} and Kamiokande~\cite{rn_sk}. The neutron flux in the
experimental hall is continuously measured with two one-liter
BC501A liquid scintillation detectors inside and outside of the main shield.
The estimated neutron flux in the experimental hall is
$8\times10^{-7}/\rm{cm}^2/\rm s$ for 1.5~MeV~$<E_{neutron}<$~6.0~MeV
which is much lower than that in the
Cheongpyung Underground Laboratory~\cite{kims_hjkim2004}.

\section{Reduction of the internal background in \csitl crystal}

The major radioisotopes in the \csitl crystal contributing to 
the internal background are $^{137}$Cs, $^{134}$Cs and
$^{87}$Rb~\cite{kims_tykim,kims_ydkim}. $^{134}$Cs has a half-life of
2~years and has a cosmogenic origin - neutron  
capture by $^{133}$Cs. Therefore, $^{134}$Cs is unavoidable unless
the crystals are stored underground for many years. However the
signal from its decay can be easily removed by the $\gamma$-ray that
follows immediately after the $\beta$ decay. The signal is large and
beyond the energy range of our interest for WIMP search.
Additional reduction can be made by using coincidence signals from
neighboring crystals. The average contamination level of $^{134}$Cs is measured
to be 20~mBq/kg, while 1~mBq/kg can contribute only less than
0.07~counts/(keV$\cdot$kg$\cdot$day)~(CPD) at 10keV. Therefore the
$^{134}$Cs contamination is not a major problem.

$^{137}$Cs, the half-life of which is 30~years, comes from a
man-made origin, mainly due to nuclear bombs and nuclear reactors. It decays to an
excited state of $^{137}$Ba by emitting an electron with a Q value
of 514~keV, followed by $\gamma$-ray emission to the ground
state $^{137}$Ba with a half-life of 2.55~minutes. Therefore the
Compton scattering of this $\gamma$-ray as well as the $\beta$-ray can
cause background in the low energy range where WIMP signals are
expected. Simulation studies show that 1~mBq/kg can contribute
0.35~CPD background in the 10~keV region. 
We investigated the contamination process and the reduction method of
$^{137}$Cs. Most of the suspected intermediate products 
from pollucite, an ore of Cs, to CsI powder has been
measured with a low background HPGe detector installed at the Y2L.
We also investigated the processing water and conclude that
the main source of
$^{137}$Cs is the processing water used for the powder production,
in which the contamination level of $^{137}$Cs was found to be
0.1~mBq/$\ell$. By using purified water, we succeeded in reducing the
$^{137}$Cs in the final powder and the crystal~\cite{kims_ydkim}.

Rb, which includes 27.8\% of $^{87}$Rb, 
exists in the pollucite at a 0.7\% level and can contaminate easily
because it is chemically similar to Cs. The $^{87}$Rb undergoes
$\beta$ decay to the ground state of its daughter nucleus, $^{87}$Sr, with
emission of an electron whose end point energy is 282~keV. This can be
a very serious background, and 1~ppb contamination can contribute 
1.07~CPD at 10keV. Contamination of many CsI powders in the
market has been measured with the ICP-MASS method and we found that it
varies from
1~ppb to 1000~ppb. 
The reduction technique of Rb, a repeated recrystallization process,
has been widely known. 
We performed the recrystallization to
the powder that we obtained with pure water and reduced the Rb
contamination
below 1~ppb. The reduction of the internal background is
discussed in detail elsewhere~\cite{kims_background}.

\section{Experimental setup}

We have installed a shielding structure in the experimental hall to
stop the external background originating mainly from the surrounding rocks.
The shield consists of 10~cm thick Oxygen Free High
Conductivity~(OFHC) copper, 5~cm thick polyethylene~(PE), 15~cm thick
Boliden lead and 30~cm mineral oil~(liquid parafin) from inside out.
The mineral oil is mixed with 5\% of a pseudocumene-based liquid
scintillator and mounted with PMTs so that it can perform as a muon
detector~\cite{kims_zhu}.
Inside the copper chamber, N$_2$ gas is flown at  a rate of 4~$\ell$/min to
reduce the radon contamination as well as to keep the humidity
low. 

The \csitl crystal~(full-size crystal) used for the experiment has a dimension of
8~x~8~x~23~cm$^3$ and a mass of 6.6~kg. The crystal is attached with two
low background quartz window PMTs with RbCs photocathode.
RbCs photocathode
enhances quantum efficiency in the green wavelength region and gives 50\%
more photoelectron yield for \csitl crystal than a normal bialkali
PMT does. As a result, the number of photoelectrons is about 5.5/keV
for the full-size crystal. 

\begin{figure}[ht]
\center \psfig{figure=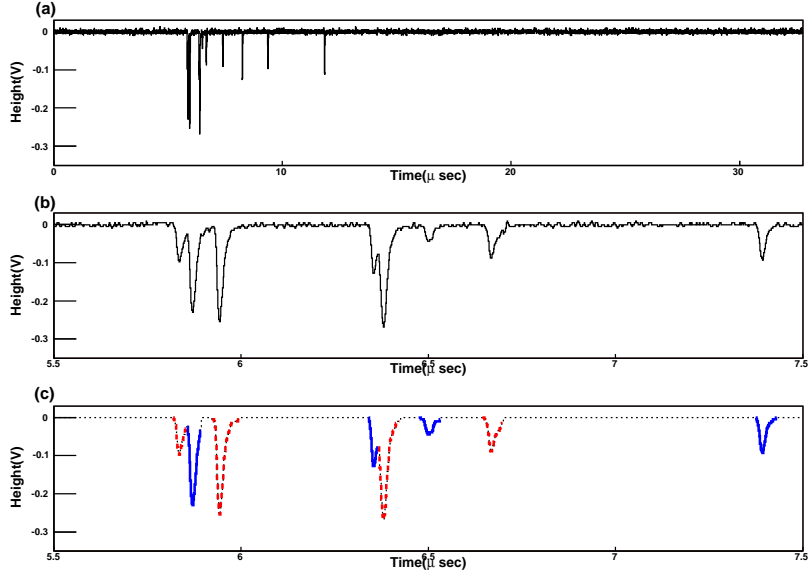, width = 5.0 in}
\caption{(a) shows typical low energy $\gamma$ signal from \csitl crystal
for one PMT obtained by Compton events from \cs
source. Zoomed pulse shape of this event from 5.5~$\mu$s to 7.5~$\mu$s
is shown at (b). The same pulse spectrum with clustering is
shown in (c). The neighboring cluster is separated by a different
style (solid line and dashed line). 
} \label{low_pulse}
\end{figure}

The signal from the PMT is amplified with a preamplifier mounted
outside the main shield and brought to the FADC module through a 20~m-long 
coaxial cable. The homemade FADC module is designed to sample the
pulse every 2~ns for a duration up to 32~$\mu$s so that one can
fully reconstruct each photoelectron pulse as shown in
Fig.~\ref{low_pulse}. The trigger is formed in the FPGA chip on the
FADC board. 
For low energy events, it is required to have more than five
photoelectrons in two~$\mu$s for the event trigger. 
An additional trigger is
generated if the width of the pulse is longer than 200~ns for high
energy events where many single photon signals are merged into a big
pulse. The FADC located in a VME crate is read out by a
Linux-operating PC through the VME-USB2 interface with a maximum data
transfer rate of 10~Mbytes/s. The DAQ system is based on
the ROOT~\cite{root} package. 


During the two-month period starting from July 2004, we
have taken data for WIMP search using one crystal with  
a background level of approximately 7~CPD at 10keV.
The amount of data was 237~kg$\cdot$days.

\section{Calibration data}

The different timing characteristics between nuclear recoil and
electron recoil in \csitl crystal make it possible to statistically
separate the nuclear recoil events from the $\gamma$
background using the mean time distribution~\cite{kims_hspark,pecourt,kudryavtsev}.
In order to have a good reference distribution of the mean time for
them, we took
calibration data of electron recoil from the $\gamma$ source and nuclear
recoil from the neutron source.

The $\gamma$ calibration data was obtained with
a full-size crystal by a $^{137}$Cs source in the copper chamber of Y2L.
Identical setup and conditions as for the WIMP search data
were used. For one week irradiation,
we took low energy $\gamma$-ray events equivalent to
approximately 3000~kg$\cdot$days WIMP search data. 

Neutron calibration data were obtained by exposing a small-size test
crystal (3~x~3~x~3~cm$^3$) to neutrons from 
300~mCi Am-Be source, prepared at Seoul National
University~\cite{kims_jiklee}. In order to
identify neutrons scattered from CsI, we used neutron detectors, 
made of BC501A contained
in a cylindrical stainless steel vessel.
Each neutron detector is shielded by 5~cm lead and 10~cm paraffin, and
set up  at various angles with respect to the incident neutron
direction.
The Am-Be source is surrounded by liquid scintillator~(LSC)
composed of 95\% mineral oil and 5\% of pseudocumene
with a collimation hole to the direction of the \csitl crystal. The LSC
acts as a tagging detector of 4.4~MeV $\gamma$'s which are
simultaneously generated with neutrons from the Am-Be source
as well as a neutron shield for the low surface neutron flux outside of
the source.

In order to identify neutron-induced events, we required a
coincidence between any one of the neutron detectors and the \csitl
crystal. With a good neutron separation capability, 
we took neutron data, whose amount depended on energy, equivalent to
approximately 1200~kg$\cdot$days WIMP search data at 3-11~keV.

We also took electron recoil data using a $^{137}$Cs source for the
test crystal used for the neutron calibration. This data is compared
with the electron recoil calibration data obtained for the full-size
crystal to confirm that neutron calibration data can be used for the
full-size crystal.

\section{Data analysis}

Single photo-electrons~(SPEs) in an event is identified by applying a
clustering algorithm to the FADC data. 
The energy deposition is evaluated from the sum of charges of all SPEs in
the event. Also, using the time information
of SPEs we calculate Mean Time~(MT). The MT distribution of events above
3~keV and up to 11~keV is used to extract the fraction of nuclear
recoil events for WIMP search. 

As one can see in Fig.~\ref{low_pulse} (b), the single clusters of low
energy events are  well reconstructed in our DAQ system.
A clustering algorithm to identify each SPE 
signal is applied for the data analysis. 
The clustering algorithm includes the identification of local maximum
to form isolated cluster using the FADC bins above the pedestal and
the separation of neighboring cluster in the case two local maximum is
found in a cluster. 
A threshold is applied to the pulse
height to select SPE candidate. Additionally the SPE candidate with an
unusually narrow pulse width out of 3$\sigma$ is rejected. 
Fig.~\ref{low_pulse} (c) shows result of clustering of (b). 
The sum of the single cluster charges for whole window~(32~$\mu$s) are
used to calculate the deposited energy. Low energy calibration is done
using a 59.5~keV $\gamma$ peak from $^{241}$Am.

\begin{figure}[ht]
\center \psfig{figure=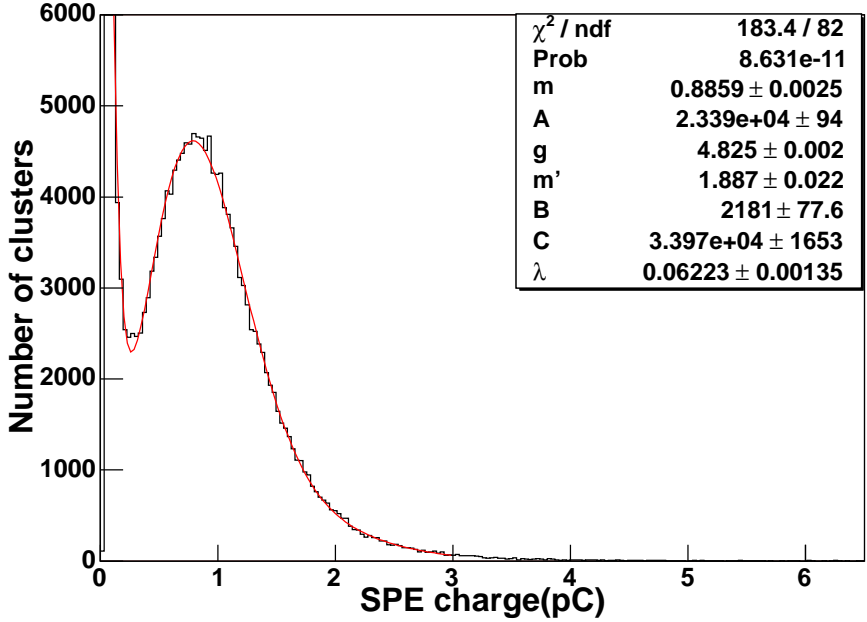, width = 5.0 in}
\caption{Single cluster charge spectrum. The distribution is
fitted with two Poisson functions. 
} \label{spespec}
\end{figure}

Fig.~\ref{spespec} shows the charge distribution of single cluster after
clustering of 5.9~keV peak from a $^{55}$Fe source without height
threshold. 
The distribution is fitted by two superimposed Poisson functions
(one for the SPE and the other for the SPE-overlapped signal) with
exponential noise component:
$$f = A \frac{\mu ^{r}
e^{-\mu}}{\Gamma (r+1)} + B \frac{\mu '^{r}e^{-\mu '}}{\Gamma
(r+1)} + Ce^{-x/\lambda}$$
$$r=xg,~\mu=mg,~\mu'=m'g$$
where $m(m')$ is
the mean of the Poisson distribution, and $g$ is the gain factor
of the PMT. The fitting function is overlaid in the figure as a
solid line. 
The ratio of the contribution of the two Poisson distributions is
9.3\%. Therefore, we conclude that $\sim$90\% of SPEs make single
clusters and $\sim$10\% makes overlapped clusters. 
The ratio of the mean values of the two Poisson
distributions, $m'/m$, is 2.11~$\pm$~0.03 which is consisted with a
expectation considering overlap of up to 3-SPEs.
 In this fit,
the most probable value~(MPV) of the SPE is obtained as 0.86~pC. From the total
charge of 59.5~keV from a $^{241}$Am source, the photoelectron yield of this
crystal is obtained as 5.5/keV.

The photoelectron yield is calibrated at the beginning and at the end
of a run. The results show stability of the light output within 1\%
for the whole
period. The time-dependent gain variation is corrected by the
MPV which is obtained from the SPE charge spectrum of low energy
(4-8~keV) WIMP search data and its Poisson fit.
Every one week's WIMP search data are accumulated to
the fit. The result shows that the gain is stable within 5\% for each
PMT in the whole period. 

\begin{figure}[ht]
\center \psfig{figure=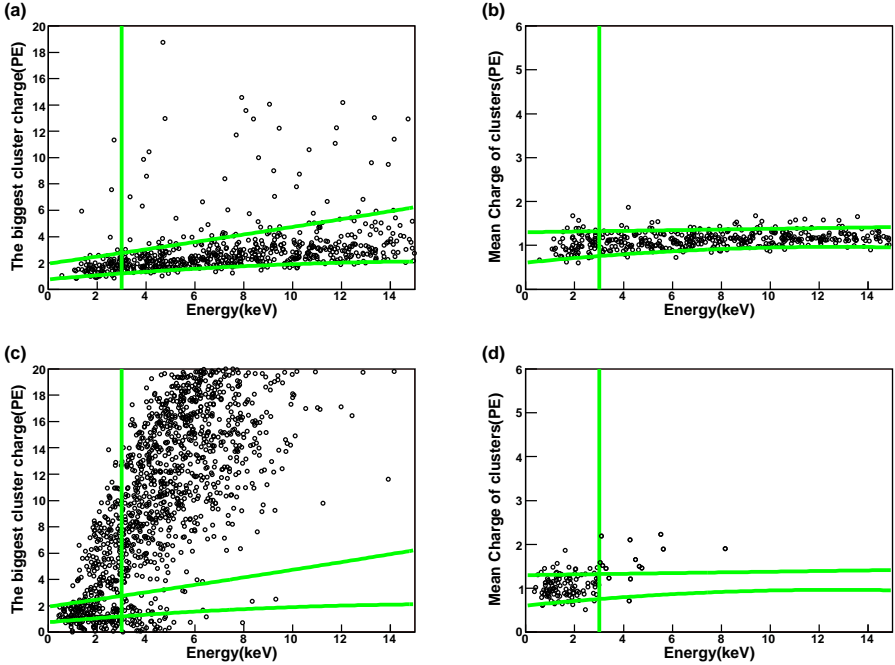, width = 5.0 in} \caption{(a)
shows the charge of the biggest cluster normalized by the MPV
of the SPE charge obtained from Fig.~\ref{spespec} fit, versus the
measured energy for the calibration data with a \cs $\gamma$ source.
Two solid lines indicate -1.65$\sigma$(lower solid line)
and 1.28$\sigma$ band~(upper solid line). The vertical line is the
3keV analysis threshold. (b) is a similar spectrum for the mean charge
of clusters for the events within the signal band
of the biggest cluster cut.  (c) and (d) are the corresponding plots
for PMT noise events.
} \label{cut}
\end{figure}

\begin{figure}[ht]
\center \psfig{figure=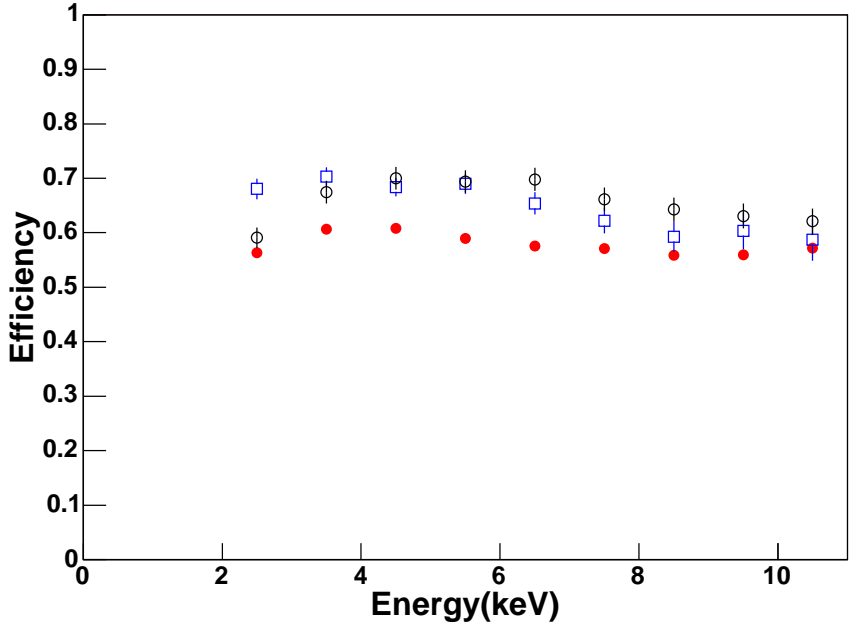, width = 5.0 in}
\caption{Efficiency calculated with $\gamma$ calibration data for
full-size crystal~(filled
circles), test crystal~(open circles), and neutron calibration data for
test crystal~(open square) where only statistical errors are included.} \label{effi}
\end{figure}

The PMT noise, which was also detected without the crystal, is seen by
both PMTs which have very fast timing characteristics. A similar
noise was reported by another group~\cite{NAIAD}. It seems
to be originated from a spark in the dynode structure~\cite{PMT}. The
PMT noise usually induces an abnormally big cluster.
Therefore, the charge of the biggest cluster 
and the mean charge of clusters for each event can be used to reject
these events.  We construct a good event-band using the Compton
scattering events from $^{137}$Cs, and they are compared with PMT
noise events in Fig.~\ref{cut}. 
PMT noise events were taken from the same system
without the \csitl crystal in the copper chamber.
The distance between the two PMTs is maintained equal to that
for the crystal-attached set-up. 
In the 25.4~days data, equivalent to
167~kg$\cdot$days WIMP search data, only two events passed all the cuts
with the 3keV energy threshold. We conclude that the PMT background
after the cuts is negligible. 
The same cut is applied to the Compton scattering data and the
efficiency was found to be approximately 60\% independent of energy as
shown in Fig.~\ref{effi}.

\begin{figure}[ht]
\center \psfig{figure=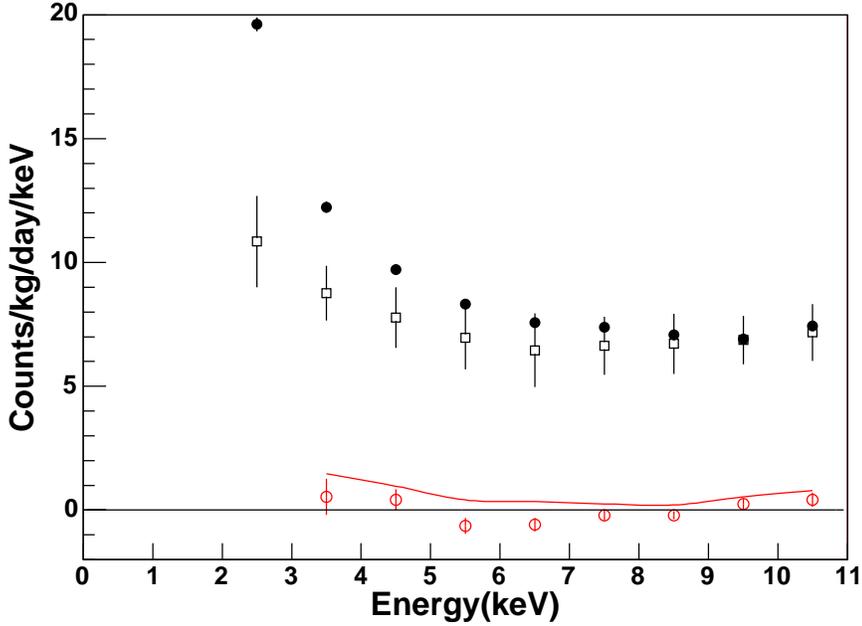, width = 5.0 in}
\caption{Energy spectrum in WIMP signal region before applying cuts~(filled circles), 
the big cluster events rejection with efficiency correction~(open
squares),
and fitted nuclear recoil rate~(open circles) where the errors include systematic
uncertainty of efficiency for the latter two cases. 
A 90\% upper limit on the nuclear recoil rate is shown with a solid
line.} \label{energy}
\end{figure}

We applied the same cuts to the calibration data taken with
the test crystal.
The efficiency of $^{137}$Cs and the neutron
calibration data for the test crystal are compared with  $^{137}$Cs data
for the full-size crystal in the Fig.~\ref{effi}. About a 10\%
difference between the full-size and  test crystals is observed.
This is mainly due to the different PMTs used. We assign a systematic
uncertainty in efficiency to account for this difference. With a similarity of
efficiency for neutron and the $\gamma$ calibration data in test
crystal, we can conclude that we can use $\gamma$ calibration data
for efficiency calculations of WIMP search data. A slight difference
in
efficiency between $\gamma$ and neutron data is also included as
a systematic uncertainty. The systematic error for the efficiency
calculation is summed  as
$$\sigma^2_{sys} = \sigma^2_{crystal~diff} + \sigma^2_{recoil~diff}$$
where $\sigma_{crystal~diff}$ is efficiency difference between
the full-size crystal and the test crystal, $\sigma_{recoil~diff}$ is
the efficiency
difference between the nuclear recoil and the $\gamma$ recoil events.
Efficiency corrected energy spectrum of events before and after the cuts
are shown in Fig.~\ref{energy}. Events below 11~keV are used
for the WIMP search.

To estimate the WIMP signal
fraction in the WIMP search data, we introduce a mean time~(MT) value which is defined as
$$<t> = \frac{\sum A_{i}t_{i}}{\sum A_{i}}-t_{0}$$
where $A_{i}$ and $t_{i}$ are the charge and the time of the $i$th
cluster respectively, and $t_{0}$ is the time of the first cluster
(assumed as time zero).

\begin{figure}[ht]
\center \psfig{figure=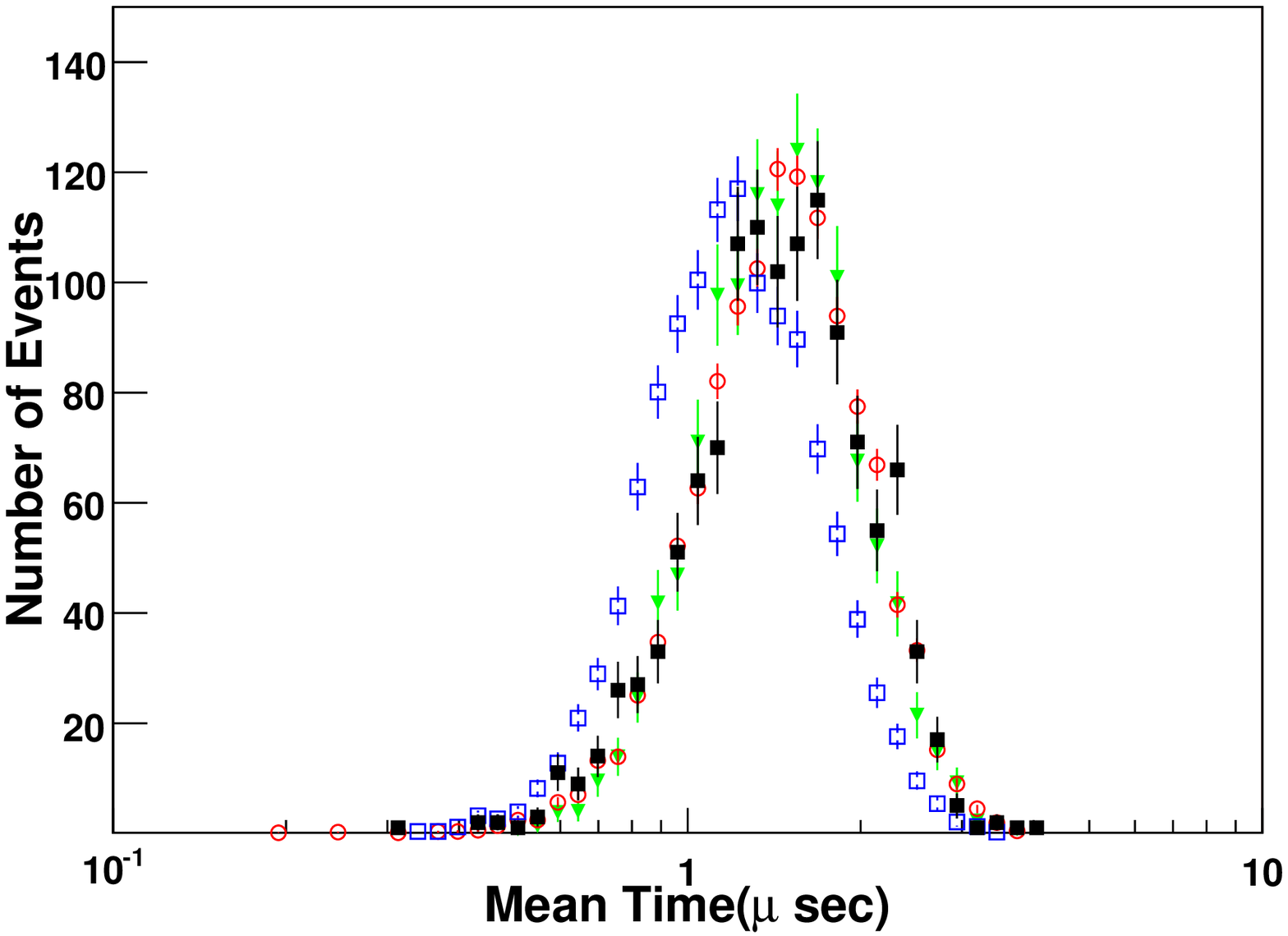, width = 5.0 in}
\caption{Mean time distributions of Compton electrons
for the test crystal~(filled triangles)
and full-size crystal~(open circles) in the 4-5~keV energy range are
compared.  Also, we include nuclear
recoil(open squares), and
the WIMP search data(filled squares) for comparison.
} \label{mtcompcry}
\end{figure}

\begin{figure}[ht]
\center \psfig{figure=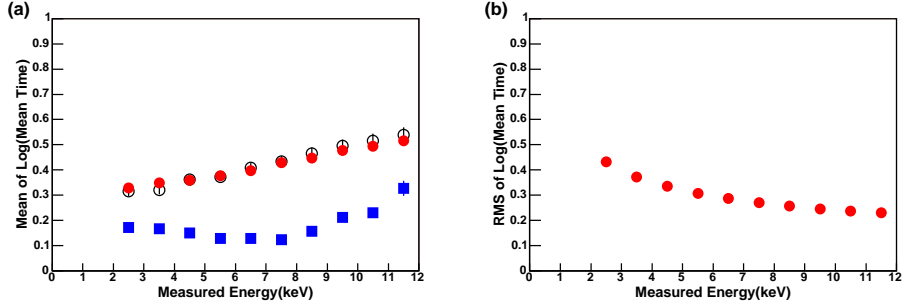, width = 5.0 in}
\caption{(a) The mean value of log(MT) as a function of measured
energy for Compton electrons with the test crystal~(open circles),
with the full-size crystal~(filled circles), and for the nuclear
recoil with the test
crystal~(filled squares). (b) Root
Mean Square~(RMS) of log(MT) as a function of measured energy for
Compton electrons with the full-size crystal. } \label{mtcompv}
\end{figure}

Because we use different crystals for the neutron calibration, we
need to confirm whether the two different crystals show the same MT
characteristics. The $^{137}$Cs calibration data of the test crystal is
compared with that of the full-size crystal. As one can see in
Fig.~\ref{mtcompcry}, the MT distribution of Compton electrons in
the test crystal is well matched with that of the full-size
crystal. The mean value of the log(MT) distribution as a function of
energy is shown in Fig~\ref{mtcompv}. An excellent agreement
between the test crystal and the full-size crystal for the Compton
electron allows us to use the neutron signal from the test
crystal as a reference for nuclear recoil signal for the full-size
crystal. A slight MT difference is adjusted by the assumption of
a constant $R_{\tau}=\tau_{n}/\tau_{e}$~\cite{NAIAD}. Where $\tau_{n}$
is the MT of
the nuclear recoil and $\tau_{e}$ is the MT of the electron recoil.

Since the MT distribution depends significantly on the measured energy
in the low energy region, the log(MT) distribution in each
keV energy bin is fitted to the reference distribution
for the same energy bin. 
The fitted
nuclear recoil event rate after the efficiency correction is given
in Fig.~\ref{energy}. The fitted nuclear recoil event rates are
consistent with zero within one standard deviation error for
all energy bins. A 90\% confidence level~(CL) upper limit on
nuclear recoil event rates are shown with a solid line. Since below
3~keV the PMT background contributes significantly and the pulse shape
discrimination power is less effective, we do not use events
below 3~keV. In order to evaluate nuclear recoil energy one needs to
know the quenching factor~(QF) defined by the $\gamma$ equivalent
measured energy divided by the nuclear recoil energy. We used the QF
measured in our previous beam test~\cite{kims_hspark}. Our threshold
of 3~keV corresponds to 20~keV nuclear recoil energy.

\section{Result and discussion}

Assuming a Maxwellian dark matter velocity distribution
with a spherical halo model discussed in Ref.~\cite{smith},
the total WIMP rate is obtained as
\begin{eqnarray}
R(E_{0},E_{\infty}) & = &
\frac{k_{0}}{k_1}\int_{0}^{\infty}dE_{R}\left\{c_{1}\frac{R_0}{E_{0}r}e^{-c_{2}
E_{R}/E_{0}r} - \frac{R_0}{E_{0}r}e^{-v^2_{esc}/v^2_{0}}\right\}
\nonumber\\
R_0 & = &
{5.47}\left(\frac{GeV/c^2}{m_{\chi}}\right)\left(\frac{GeV/c^2}{m_t}\right)\left(\frac{\sigma_0}{pb} \right)
\left(\frac{\rho_{\chi}}{GeV/c^{2}/cm^{3}}\right)
\left(\frac{v_0}{km/s}\right)
\nonumber\\ 
E_0 & = & \frac{1}{2} m_{\chi} v_{0}^{2} \qquad r  =  \frac{4m_{\chi}m_t}{(m_{\chi}+m_t)^2}
\label{tru}
\end{eqnarray}
where $R_0$ is the event rate per kg$\cdot$day for $v_E =0$ and
$v_{esc}=\infty$, $v_{esc}=650$~km/sec is the local Galactic
escape velocity of WIMP, $m_t$ is the mass of
a target nucleus, $\rho_\chi=0.3$GeV/cm$^3$ is local dark matter density, 
$v_0 =220$~km/sec is a Maxwell velocity parameter, and $c_1$, $c_2$ are
constants, as discussed in Ref.~\cite{smith}. 

\begin{figure}[ht]
\center \psfig{figure=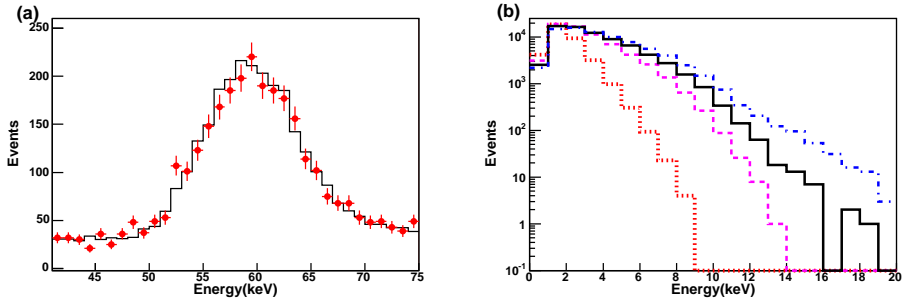, width = 5.0 in}
\caption{(a) Distribution of $E_{ee}$ for the MC simulation~(solid
line) and calibration data~(filled circles) for a $^{241}$Am source are
compared. (b) Simulated 
$E_{ee}$ spectra for several WIMP masses (20~MeV - dotted
line, 50~MeV - dashed line, 100~MeV - solid line, 1000~MeV - dotted
dashed line) are shown.}
\label{Wimp_simu}
\end{figure}

In order to estimate the expected event rates for each energy bin,
we use MC simulation. The MC simulation based on
GEANT4~\cite{geant4} takes into account the recoil energy
spectrum, the QF, and the light transportation to the PMTs. Then the simulated events
are analyzed in the same way as the data except for the applying
any analysis cuts. The energy is tuned to provide good agreement
with the calibration data using 59.5~keV $\gamma$-rays from a $^{241}$Am
source. Fig.~\ref{Wimp_simu}~(a) shows good agreement between Monte
Carlo and calibration data for the energy distribution. 
The Monte Carlo generated electron equivalent energy distributions
($E_{ee}$) for several
WIMP masses are shown in Fig.~\ref{Wimp_simu}~(b).

From the rate of nuclear recoil in each energy bin, we can estimate
the total WIMP rate in comparison with the
simulated $E_{ee}$ distribution for each WIMP mass by the following relation.
\begin{equation}
R(E_{0},E_{\infty}) = R_{E_{k}}N_{total}/N_{E_{k}}
\label{Rcal}
\end{equation}
where $R_{E_{k}}$ and $N_{E_{k}}$ are the measured nuclear recoil rate and
the simulated WIMP events for each energy bin $E_{k}$ respectively and,
$N_{total}$ is the total number of WIMP events generated by simulation. With
Eq.~(\ref{tru}) and Eq.~(\ref{Rcal}), 
we can convert the rate of nuclear recoil in each energy bin to the
WIMP-nucleus cross section for each WIMP mass.

\begin{figure}[ht]
\center \psfig{figure=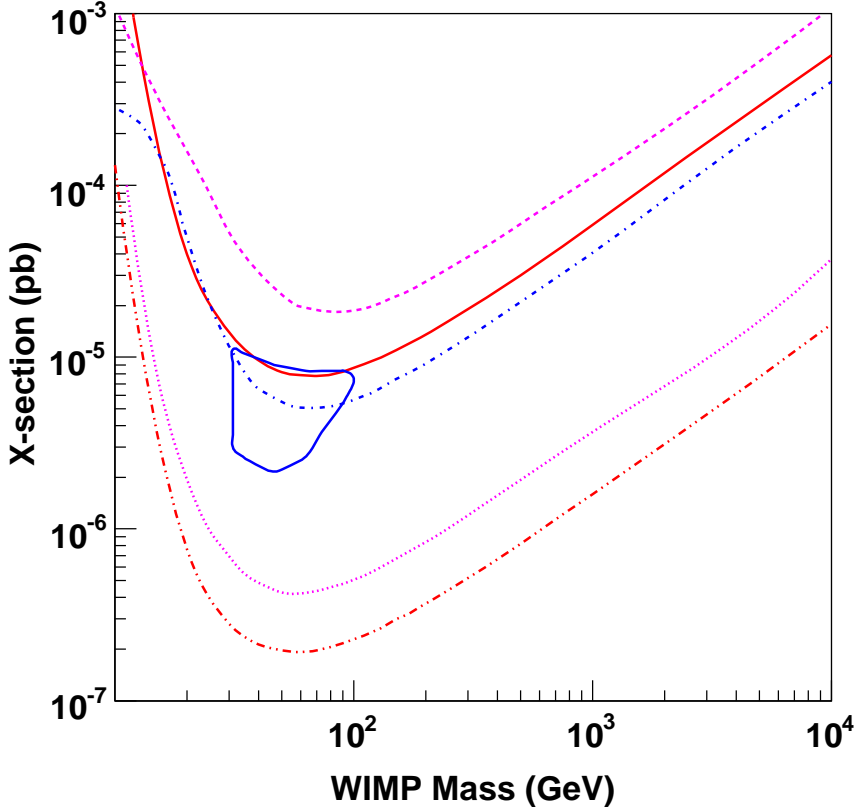, width = 5.0 in}
\caption{The KIMS limit on a WIMP-nucleon cross section for a spin
independent interaction with 237~kg$\cdot$days exposure (solid line),
DAMA positive~\cite{DAMA} annual modulation signal (closed curve), NAIAD
limit~\cite{NAIAD}
with 3879~kg$\cdot$days exposure (dashed line) and, DAMA
limit~\cite{DAMA96} with 4123 kg$\cdot$days (dashed-dotted line) are
presented. Also the KIMS projected limit with 250~kg$\cdot$year exposure of 2~CPD
background level(dashed-double-dotted line) is presented. 
Dotted line shows current best limit by CDMS group~\cite{CDMS}.} \label{1stlimit}
\end{figure}

The limits on the cross-section for various energy bins and targets (Cs
and I) have been combined following the procedure described in
Ref.~\cite{smith} assuming the measurements for different energy bins
are statistically independent. 
The combined result from energy bins for a WIMP-nucleus
cross section is obtained from this expression
\begin{eqnarray}
\sigma_{W-A} & = & \frac{\sum
\sigma_{W-A}(E_{k})/\delta\sigma^{2}_{W-A}(E_{k})}{\sum
1/\delta\sigma^{2}_{W-A}(E_{k})}
\nonumber\\
\frac{1}{\delta\sigma^{2}_{W-A}}
& = & \sum \frac{1}{\delta\sigma^{2}_{W-A}(E_{k})}
\label{sum}
\end{eqnarray}
where $\sigma_{W-A}$ is a combined WIMP-nucleus cross section,
$\sigma_{W-A}(E_{k})$ is a WIMP-nucleus cross section calculated in
an energy bin $E_{k}$. As one can see in Fig.~\ref{energy}, the rate
of nuclear recoil events is consistent with zero. Therefore, we
can set the 90\% CL upper limit on the WIMP-nucleus cross section with
Eq.~(\ref{sum}). In this process, we assign zero as the mean value
for the event rate for the bins with negative means. 
The WIMP-nucleon cross section can be obtained from
WIMP-nucleus cross section by following equation
\begin{equation}
\sigma_{W-n}=\sigma_{W-A} \frac{\mu_n^2}{\mu_A^2}
\frac{C_n}{C_A}
\end{equation}
where $\mu_{n,A}$ are the reduced masses of WIMP-nucleon and
WIMP-target nucleus of mass number A and $C_A/C_n = A^2$ for spin
independent interaction.
The limit on the WIMP-nucleon cross section for each
nucleus can be combined by this expression
\begin{equation}
\frac{1}{\sigma} = \frac{1}{\sigma_{Cs}} + \frac{1}{\sigma_{I}}
\end{equation}
A 90\% CL upper limit on the WIMP-nucleon cross section from CsI for spin
independent interaction is shown in Fig.~\ref{1stlimit}, together with the
limits obtained from two \naitl crystal based WIMP search experiments with similar pulse
shape analyses, NAIAD(UKDMC)~\cite{NAIAD} and DAMA~\cite{DAMA96}.
Although the amount of data used to get our limit is 10 times less than
that of NAIAD, we achieved a more stringent limit than that of NAIAD due to the 
better pulse shape discrimination and lower recoil energy
threshold.

\section{Conclusion}

The KIMS collaboration has developed a low background \csitl
crystal for the WIMP search. We set the first limit on the WIMP cross
section using the 237~kg$\cdot$days data taken with a 6.6~kg crystal.
Our limit already partially excludes the DAMA 3$\sigma$ signal
region. The current experimental setup is designed to accommodate
about 250~kg of \csitl crystals without any modification. We
expect that the background rate will be reduced to the level of
approximately 2~CPD or less
for the new powder produced with purer water. The projected limit for
1 year of data taken with 250~kg crystals of 2~CPD
background is shown in Fig.~\ref{1stlimit} in comparison with the
current best limit set by CDMS~\cite{CDMS}. With the 250~kg setup,
KIMS can explore the annual modulation as well.

\section{Acknkowledgment}
\label{}

This work is supported by the Creative Research Initiative program
of the Korea Science and Engineering Foundation. We are grateful to
the Korea Middleland Power Co. and the staff members of the YangYang Pumped
Power Plant for their providing us the underground laboratory
space.

\end{document}